\documentstyle[11pt,aaspp4]{article}

\begin{document}
\title{MORPHOLOGY OF COLD BARS \\
IN EARLY AND LATE TYPE GALAXIES}

\author{
A. C. Quillen\altaffilmark{1}$^,$\altaffilmark{2}
}

\altaffiltext{1}{Astronomy Department, Ohio State University, 174 W. 18th Ave., 
    Columbus, OH 43210}
\altaffiltext{2}{E-mail: quillen@payne.mps.ohio-state.edu}

\begin{abstract}
We compare stellar orbits in an early (NGC~4314) and a late-type barred galaxy 
(NGC~1073).  We find that these bars are cold in the sense that the majority of 
stars can be described as being nearby to periodic orbits.
We place limits of (65km/s)$^2$ and (50km/s)$^2$ on the components of
the diagonalized velocity dispersion ellipsoid matrix for stars in the
bars of NGC~4314 and NGC~1073 respectively.
Both bars end near the inner 4:1 Lindblad (ultraharmonic) resonance.  
We conjecture that
a large class of bars end near this resonance.  
The morphology of the bars depends
on the strength of the 4:1 resonance, which is strong in the early-type
barred galaxy and weak in the other.
This results in a flat bar major axis surface brightness 
profile for the early-type bar
and an exponential profile for the late-type bar.

\end{abstract}


\section {Introduction}

Many studies of the dynamics of stars in model barred galaxies have
concentrated on exploring the structure of periodic orbit families 
(e.g. \cite{ath92a}) and the ergodic structure of resonances (\cite{con89}).
However, little is known about the actual distribution function of stars
in real barred galaxies, or how stars are distributed among the different
orbits.  
One approach is to calculate a library of orbits and iteratively construct 
a self consistent galaxy model using statistical methods
(the Schwarzschild method; \cite{sch79}). 
This approach yields many solutions and has primarily
been used to explore the properties of model galaxies 
(most notably by \cite{pfe84}).
Fine structure observed in the images of barred galaxies 
is not usually  present in model or N-body galaxies.  This fine structure 
can exist only in a system with a sufficiently low stellar velocity dispersion.
In this paper we place limits on the velocity dispersion 
in two barred galaxies of different morphology and Hubble type
by comparing the properties of their orbits with their shapes.

Bars in barred spiral galaxies have been observed to have major
axis surface brightness profiles of two forms, either flat
or dropping exponentially with increasing distance from the nucleus 
(\cite{elm85}).   These two types of barred
galaxies can be called exponential-type bars and flat-type bars.
Exponential-type bars tend to be in later type galaxies than
flat-type bars (\cite{elm96a}).
Although there have been a few studies 
of orbits in individual barred galaxies based upon observations of these galaxies
(e.g. \cite{qui94} \& \cite{ken89}),
there have been no studies comparing the properties of orbits in different galaxies.
In this paper we find the periodic orbits (closed orbits in the frame
in which the bar is stationary) in the plane of
the galaxy integrated in the gravitational
potential of the exponential and late-type barred galaxy NGC~1073. 
This paper is a continuation of \cite{qui94} (hereafter Paper I) where periodic
orbits were found for the flat and early-type barred galaxy NGC~4314.
The bar pattern speeds and the disk vertical scale heights are measured.
The morphology of the two galaxies and their orbit shapes are compared.
We account for the difference in morphology of flat-type and 
exponential-type bars based on this orbit analysis.
We also explore orbits that are nearby the periodic orbits in 
both galaxies to investigate the effect of the stellar velocity dispersion.

\section{The $J$ band image of NGC~1073}
NGC~1073 was chosen from the exponential-type bars listed in \cite{elm85}.  
These data are a preliminary part of a survey being carried
out at Ohio State University of $200$ to $300$
galaxies that will produce a library of photometrically calibrated
images of late-type galaxies from $0.4$ to $2.2 \mu$m.
For notes on the reduction and observation techniques see
\cite{pog96}, or for individual examples see Paper I, and \cite{qui95}.
The galaxy was observed at
the $1.8$m Perkins Telescope of the Ohio State and Ohio Wesleyan
Universities at Lowell Observatory in Flagstaff, AZ using a $256 \times
256$ HgCdTe array in the Ohio State Infra-Red Imaging System (OSIRIS)
(\cite{dep93}).   The scale is $1.50$ arcsec/pixel.  

The $J$ image (see Figure 3b) 
was used for this study because it has the highest 
signal to noise of the near infrared images
($J$, $H$, and $K$) included in the survey.  
A $J-H$ color map reveals no detectable 
color changes (less than 0.03 mag)
across the galaxy suggesting that the mass-to-light 
ratio is approximately constant across the bar and that the $J$ image
is not strongly affected by extinction from star formation or dust.
As discussed in Paper I near infrared images are superior to
visible images for dynamical studies because of their reduced
sensitivity to extinction from dust and because they are dominated by
light from an older cooler stellar population that is more
evenly distributed dynamically and a better 
tracer of the stellar mass in the galaxy than the bluer,
hotter stars (e.g. \cite{fro88}).

\section{Morphology of the galaxy as compared to NGC~4314 }

We decompose the surface brightness at $J$ band 
(and subsequently the potential)
into its Fourier components in concentric annuli (e.g. \cite{elm89}). 
In Figure 1 we show these components given by
$S(r,\theta) = S_0(r) + \sum_{m>0} S_{mc}(r) \cos(m \theta)
+ S_{ms}(r) \sin(m \theta)$ where $S(r,\theta)$ is the galaxy
surface brightness.
Fourier components of the potential in the plane of
the galaxy, similarly defined, are shown in Figure 2.
NGC~4314 has a stronger bar than NGC~1073 in the sense that
the maximum of $S_2/S_0$ reaches 0.9 in NGC~4314 but only 0.7 in
NGC~1073.

The knobby features (density enhancements) at the ends of the
bar in NGC~4314 can be seen as strong higher order magnitude
($S_4$ and $S_6$) Fourier components that peak near the knobby features.
In NGC~4314 the $S_4$ and $S_6$ maxima are 0.7 and 0.5 times 
the $S_2$ maximum respectively, whereas in NGC~1073 
the $S_4$ and $S_6$ maxima are only 0.5 and 0.3 times the  $S_2$ maximum.
Because the $S_2$ moment dominates in NGC~1073, the bar has
a more elliptical appearance.
In NGC~4314 the central $20''$ containing the bulge is almost round whereas
the isophotes of NGC~1073 are elliptical even near the nucleus
(see contour plots in Figure 3).

\section{Estimating the potential}

The gravitational potential in the plane of the galaxy was derived by 
convolving the $J$ image of NGC~1073 with a function
that depends on the vertical structure of the disk (Paper I).
Before convolution stars were removed from the $J$ image.  The galaxy
is assumed to be face on, since the axis ratio at large radii ($\sim 0.9$)
is close to one.  The disk is assumed to have density 
$\propto {\rm sech}(z/h)$ (\cite{vdk88}) where $z$ is the height 
above the plane of the 
galaxy following \cite{vdk88}. The resulting potential
is insensitive to the choice of vertical function 
for functions such as sech, sech$^2$ and exponential with equivalent 
$\langle z^2 \rangle$ (\cite{qui96}).  

Periodic orbits in NGC~1073 were found by numerical integration as in 
Paper I.
The orbits in the plane of the galaxy were integrated in a potential 
resulting from  
smooth polynomial fits to the even order moments of the potential.
Components of the potential are shown in
Figure 2 along with the polynomial fits to them.
The curvature of $\Phi_0$, the azimuthally averaged component of
the potential, is far higher in NGC~4314 than NGC~1073 because of its bulge.
NGC~1073 has a weaker $\Phi_2/\Phi_0$ maximum 
than NGC~4314 and the higher order 
Fourier maxima of NGC~1073 are also weaker than those of NGC~4314.
This is not surprising since the higher order Fourier components  of the
surface brightness are also weaker in NGC~1073.

\section{Bar pattern speed and vertical scale height}

Periodic orbits found in NGC~1073 
are displayed with a contour plot of the galaxy in Figures 3a and 3b.
For comparison Figures 3c and 3d shows similar results
for NGC~4314 from Paper I.
We display a rotation curve and the location of the resonances for NGC~1073 
in Figure 4.

In Paper I we found that 
both the vertical scale height and the pattern
speed of the bar could be determined by 
comparing the periodic orbits to the shape of the galaxy in the limit
that most stars in the galaxy are in orbits that are nearly periodic.
The bar pattern speed is determined from the location of a resonance that
causes features observed in the galaxy, whereas
the vertical scale height is determined from
the ellipticity of the orbits at regions distant from resonances.
Some iteration is required to estimate both parameters since
the rotation curve which determines the location of resonances
depends on the vertical scale height, and 
the ellipticity of the orbits depends on the bar pattern speed.

We find that the orbits have ellipticities similar to the galaxy 
for a vertical scale height $h=4'' \pm 2$.   
We find a vertical scale height to exponential scale
length ratio of $\sim 1/12$  
using a disk exponential scale length of $50''$ (in $I$ band \cite{elm85}).
This ratio is consistent with infrared studies of 
edge on galaxies (Barnaby \& Thronson 1992; Wainscoat et al. 1989).

NGC~1073 has no distinct
features obviously related to a resonance such as the knobby features at 
the ends of the bar in NGC~4314 caused by the inner 4:1 Lindblad resonance,
which allowed us to fix the pattern speed of the bar.
However we find that the orbits of NGC~1073 correspond to the shape
of the galaxy only for pattern speeds that placed the 4:1 resonance
at the end of the bar.  
This is because lenticular shaped orbits that are strongly peaked along 
the bar major axis exist only within the 4:1 resonance.

For early-type, strongly-barred galaxies, corotation should be close
to the end of the bar, as predicted theoretically by
\cite{con80} and others.
Using a measurement of the bar pattern speed
various studies have estimated the ratio of the corotation radius
to the bar semi-major axis in early-type galaxies.
\cite{ken90} found that this ratio was $\sim 1.2$ and later studies
have confirmed this finding ratios of $1.2 \pm 0.2$
(\cite{ath92b}) and  $1.3 \pm 0.1$ (\cite{elm96a}).
In particular \cite{elm96a} noted that bars in early-type
galaxies ended between the 4:1 resonance and the corotation radius
but did not associate the end of the bar with any particular
resonance.  Because the 4:1 resonance is near  
the corotation radius in early-type galaxies it is difficult
to clearly associate the end of the bar with the 4:1 resonance,
though we noted in Paper I that in NGC~4314 the 4:1 resonance caused
the knobby features at the end of the bar,
and was coincident with the radius at which the isophotes began to twist.
However in late-type galaxies the distance between the 4:1 resonance
and the corotation radius is significantly larger than in early-type
galaxies because of the shallower slope of the rotation curve.   
This makes it possible
in NGC~1073 to clearly associate the end of the bar with the 4:1 resonance.
That an exponential late-type bar also ends near the 4:1 resonance
(well before corotation) is unexpected.    
We conjecture that a large class of bars end near this resonance.

The azimuthally averaged component of the potential gives a rotation
curve (see Figure 4) 
for NGC~1073 which rises slowly and peaks past the end of the bar,
not untypical for a late-type galaxy, whereas the rotation curve of NGC~4314
peaks near the end of its bar (see Figure 5 of Paper I).
We derive a maximum circular velocity of $160$ km/s assuming a distance
of 16Mpc (with a Hubble constant of $75$ km/s Mpc) and using a mass-to-light
ratio of $M/L_J=4.3$ in units of solar masses to solar $J$ magnitudes
where $L_J$ is the luminosity in the $J$ band.
Using these quantities we find a bar angular rotation rate 
or pattern speed of $\Omega_b = 0.0237 \pm 0.005 $Myr${^{-1}}$.

Our pattern speed places the corotation radius at $\sim 85''$ 
which puts the radius of corotation at a distance of 
1.7 times the bar semi-major axis, assuming that the
bar ends at $\sim 50''$.  At $\sim 50''$ the magnitude of the $S_2$ component
of the surface density experiences a drop and begins to twist (see Figure 1).
1.7 is actually a lower bound for this
ratio since dark matter, which was not taken into account in constructing the
rotation curve, will cause the radius of corotation to be at an even
larger radius.  This ratio is
significantly higher than that found for early-type bars such as 
NGC~4314 (1.0-1.4) (\cite{ath92b}; \cite{elm96a}).   
The larger distance between
the end of the bar and the corotation radius in the late-type
galaxy allows us to clearly associate the end of the bar with
the 4:1 resonance.  

We find that NGC~1073 does not have an inner Lindblad resonance 
(see Figure 4) which implies that the entire
length of the bar lacks resonances (except for a possible weak 3:1).
This causes many of the orbits within the bar to be quasiperiodic, described in 
phase space by tori about the periodic orbits shown in Figure 3.
In the next section we discuss the morphology of orbits that
are not periodic.

\section{Limiting the stellar velocity dispersion}

In the previous section we have used the resemblance of the periodic 
orbits to the shape of the galaxy to determine the pattern speed and
the scale height of the disk in the limit 
that all orbits are nearly periodic.
To investigate the role of the stellar velocity dispersion in the plane
of the galaxy we have integrated
orbits near the periodic orbits (shown in Figure 3).  
Some non-periodic orbits are 
displayed in Figure 5 for NGC~1073 and in Figure 6 for NGC~4314.
These figures show groups of orbits with identical initial positions
but with varying initial velocities.

The stellar velocity dispersion can be defined locally in terms
of a velocity dispersion ellipsoid.   \cite{pfe84} described the local 
velocity dispersion in terms of a sum of a dispersion from individual
orbits, and that from the sum of all the orbits.
By comparing Figures 5 and 6 showing the non-periodic orbits to Figure 3 
showing the galaxy isophotes we can see that 
most stars in these galaxies must be in orbits that are close to being 
periodic.
The velocity dispersion cannot be too high in NGC~4314 otherwise
the knobby features observed at the ends of the bars could not exist.
Likewise the strongly peaked top of the bar in NGC~1073 could not
be sustained if the velocity dispersion were high.
This enables us to place a limit on the velocity dispersion in both
galaxies.   
  
The second orbit from the bottom
in Figure 5a and the same in Figure 6b were used to compute velocity
dispersion ellipsoids at various locations in the orbit.  Having
a large percentage of stars in orbits near to these orbits 
would not cause the galaxy to have a shape different than observed,
however stars in orbits with larger velocity dispersions could not
be represented in large numbers otherwise the
galaxies could not have the fine structure observed.  

We computed velocity dispersion ellipsoids in wedges centered
at different position angles
by considering all points in an orbit that were within an angular
distance of 0.2 radians from the center of the wedge.
Since these points cover a range of
radius, this should be a reasonable estimate for the contribution to
the dispersion from similar orbits at smaller and larger average radius.
The largest component of the velocity dispersion ellipsoid occurs
at the apocenters of the orbit where 
the ellipsoid is aligned in the radial direction.
We find that the largest components of the 
the diagonalized velocity dispersion ellipsoid matrices are  
(65km/s)$^2$ and (50km/s)$^2$ 
in NGC~4314 and NGC~1073 respectively.

We note that 
outside the 4:1 resonance a significant number of stars could still 
be in orbits that cover a large area (e.g. \cite{ath83};
\cite{con89}).   In fact a large observed velocity
dispersion is predicted at the end of the bar (\cite{wei94}).

We notice that the higher the initial velocity 
the rounder and larger the area covered by the orbit.
Bars with larger velocity dispersion are expected to be rounder and more
oval in shape.
Because we have used the periodic orbit shapes to determine
the vertical scale height, our estimate for the scale height is biased 
and is actually an overestimate.
However we find that the size of the overestimate is
smaller than the uncertainty in our value for the scale height itself. 
Likewise the initial assumption of a vertical scale height causes our
limit on the velocity dispersion to be slightly underestimated.
We note that by not including stars on ergodic orbits we have also 
underestimated this limit.

The orbits are bounded in extent by contours of an effective potential.
In a system with a constant pattern speed the Jacobi integral given by 
$$J = \Phi - {1 \over 2}\Omega_b^2 r^2 + {1\over 2}v^2$$
is a conserved quantity.    
An effective potential can be defined as 
$$\Phi_{eff} \equiv \Phi - {1\over 2} \Omega_b^2 r^2.$$
A star with a value $J$ for the Jacobi integral is therefore
bounded by the contour $\Phi_{eff} = J$.
Orbits with a given $J$ value range 
from being periodic orbits to filling a region bounded
by a contour of constant effective potential.
In Figure 7 we show contours for the effective potential for both
galaxies.  By comparing Figures 5 and 6 to Figures 7a and 7b we can
see that stars with larger initial velocities fill regions that 
are bounded by the contours of constant effective potential.
Stars in these orbits can affect the outer isophotes of the bar causing it
to be rectangular or lenticular in shape (\cite{ath90}, \cite{ath91}).
Because the amount of time a star spends in a particular region can vary, 
the projected density of an orbit may have a boxier or more lenticular shape than the 
area covered by the orbit (see Figures 5 and 6).

The contours of the effective potential are quite different
for NGC~4314 and NGC~1073 (see Figure 7).  The bar in NGC~4314 forms a deeper
well since this galaxy is of an earlier type.
Near corotation the two terms ${1\over 2} \Omega_b r^2$ and
$\Phi$ are about the same magnitude.   In NGC~1073, since corotation
is located far outside the bar, the effective potential
is rounder near corotation than in NGC~4314 and is affected by
spiral arms.
In NGC~4314 stars past the end of the bar or with a high velocity 
dispersion are 
are trapped within the deep contours of the effective potential and
so can support the shape of the bar.  
In NGC~1073, however, stars past the bar or with a high velocity
dispersion will not support the bar, but will cover a more circular
region.

\section{The major axis surface brightness profile}

It is possible to construct 
the surface density image of a galaxy from the periodic orbits 
if its brightness profile along some axis is known.
Consider an initially exponential disk in which a bar slowly forms.  
Stars initially on orbits that are approximately circular
end up in orbits that are close to periodic orbits such as shown in Figure 3.  
The azimuthally averaged surface brightness for both NGC~4314 
and NGC~1073 are approximately exponential and so could be 
consistent with these bars having
formed from initially exponential disks.

For NGC~1073 the orbits are similar in shape.  Distributing stars
on periodic orbits with surface density decreasing exponentially 
as a function of radius
causes the resulting galaxy to have exponential surface brightness profiles 
along all position angles (including the major and minor axes of the bar).
However in NGC~4314 the periodic orbits become increasingly peaked along the bar
near the 4:1 resonance.  Orbits near the 4:1 resonance have much lower
speeds at the ends of the bar which results in a higher surface brightness 
than that away from this resonance.
An initially exponential disk would have 
a strong density enhancement with respect to the initial disk 
along the top of the bar.  The strength of the density enhancement 
grows as the distance to the 4:1 resonance decreases, 
reaching a maximum near this resonance.  This results in 
a flat surface brightness profile along the bar major axis.

This effect can also be seen by considering the extreme values in the
periodic orbits within the 4:1 resonance.  In Table 1 we show axis
ratios, velocity contrasts and maximum curvatures for 
the periodic orbits shown in Figure 3. 
The axis ratio is computed  as the minimum radius divided by the maximum 
radius in the orbit ($r_{min}/r_{max}$), 
the velocity contrast is computed as the maximum speed
in the orbit divided by the minimum speed in the frame in which the bar
is stationary ($v_{min}/v_{max}$), and the curvature 
we compute the dimensionless quantity 
$|x''y' - y'' x'|r_{max}$ where $x' = dx/ds$, 
$ds$ is distance along a segment of the orbit,
and the other derivatives are defined similarly.  This curvature, which
is the inverse of the radius of curvature, is the same
quantity as measured in \cite{ath92a}.

As we can see from Table 1, the axis ratios in both galaxies
decrease as a function of radius, but rate of decrease is higher 
in NGC~1073.  \cite{ath92a} found that there is a stronger variation in 
the axis ratio as a function of radius for bars with 
large Lagrangian radii (same as low
pattern speeds), and for galaxies with larger bulges.
Although NGC~1073 has a larger Lagrange radius with respect to the end
of the bar than NGC~4314 it also has a smaller bulge.

A high velocity contrast in the orbit causes the surface
brightness along the bar to increase compared to the azimuthally averaged value.
This follows from conservation of mass in a steady state system.
NGC~1073 has a slight decrease in the velocity contrast with
increasing radius, whereas NGC~4314 has a slight increase which would
increase the surface brightness along the bar major axis.

A high curvature at the end of the orbits also increases
the surface brightness directly along the major axis of the bar. 
The curvature of the ends of the orbits in NGC~1073 
does not vary much as a function of 
bar semi-major axis, though it is quite high.
However even at a semimajor axis of $50''$, 
the curvature of the ends
of the orbits in NGC~4314 is almost twice that  
at $30''$ and at a major axis of $55''$ it is four times that at $30''$. 
So NGC~4314 has a large increase in curvature near the 4:1 resonance.
Because we also see an increase in the curvature
at the $50''$ orbit the effect of the
4:1 resonance is stronger and extends over a larger radius in NGC~4314
than in NGC~1073.
The combined effect of the increasing velocity contrast and curvature
with increasing bar semi-major axis causes the surface
brightness along the bar to be higher than the azimuthally averaged  value
and so is responsible for the flat bar profile shape and 
knobby features observed at the ends of the bar in NGC~4314 (see Figure 3d).

\section {Summary and Discussion}

In this paper we have compared orbits in the exponential late-type barred 
NGC~1073 and the earlier type flat-type barred galaxy NGC~4314.
We find that the bar in both galaxies end near the inner 4:1 Lindblad (or
ultraharmonic) resonance.
In NGC~1073 the
ratio of the radius of the end of the bar to the corotation radius
is $\sim 1.7$  significantly longer than that found for early-type bars such
as NGC~4314 (1.2-1.4) (\cite{elm96a}).  
We find no inner Lindblad resonance in NGC~1073 so that 
the length of the bar is resonance free. 
By comparing the shapes of non-periodic orbits with the appearance of the galaxy
we find that the bulk of the stars in both galaxies must be in orbits
nearby to periodic orbits of the $x_1$ family.
We place limits of (65km/s)$^2$ and (50km/s)$^2$ on the components of
the diagonalized velocity dispersion ellipsoid matrix for stars in the
bars of NGC~4314 and NGC~1073 respectively.
These low values suggests that these bars formed and evolve by non-violent
processes.

Both NGC~1073 and the early-type galaxy NGC~4314 have surface
brightness profiles along the major axis that are consistent 
with their bars forming from 
an initially exponential disk.  The strong 4:1 resonance and 
highly peaked orbits near this resonance 
in NGC~4314 cause the large density enhancement along the bar
which results in a flat bar major axis surface brightness
profile.  A weaker 4:1 resonance and 
orbits that are similar to one another in NGC~1073 result in exponential
profiles along all position angles.

That both an exponential late-type bar (NGC~1073) and early-type
bars end near the 4:1 Lindblad resonance 
is unexpected.  One possible reason for this is that asymmetric terms
in the potential cause the $x_1$ family past the 4:1 resonance to
be highly asymmetric (Patsis et al.  1996a,b).  In Paper I we noted
that spiral structure, causing NGC~4314
to be asymmetric, begins at the location of the 4:1 resonance.
The same is true in NGC~1073.
We conjecture that many strong and slowly evolving bars
end near this resonance independent of Hubble type.
Since spiral arms can begin and end near their 
inner and outer 4:1 resonances (\cite{pat94}) this may 
facilitate the driving of spiral arms by bars in early-type galaxies
(which have strong 4:1 resonances)
and possibly explain why later type barred galaxies have weaker spiral arm
patterns (\cite{elm96a}).  In early-type galaxies 
the spiral arm pattern could therefore exist
with the same pattern speed as the bar, extend inside
and outside the corotation radius and have a dust lane which
crosses the spiral arm at corotation.

The stellar velocity dispersion in the bar should depend upon how the
bar was formed.  This means that our limit for the velocity
dispersion in these galaxies could be used to constrain the
timescale for bar formation.  This timescale 
depends strongly on the dark matter content and bulge size in the galaxy 
(\cite{ath86}) so that limiting the timescale for the formation
of the bar could also be used to constrain the 
dark matter content in the central regions of galaxies.

The Schwarzschild method could be used on galaxy images 
to more carefully constrain the form of the stellar distribution
function.  This would give unbiased estimates of
the bar pattern speed and the vertical scale height as well as make
predictions for stellar velocity profiles in the bar that could  
be observed with a high resolution spectrograph.  In this paper
we have only considered orbits in the plane of the
galaxy, but using the Schwarzschild method a self consistent
three dimensional system could be studied.

We note that because of the large angular pixel scale of our images
they were spatially undersampled.  Higher resolution images should
show more structure in these bars and will more 
tightly constrain the velocity dispersion.

Both galaxies studied here appear to have a low velocity dispersion for the
bulk of the stars within the bar. This suggests that the difference in morphology
of the two bars depends on the shape of the rotation curve or the Hubble
type.  We would then expect that earlier type galaxies
have even stronger 4:1 resonances and stronger knobby features at the ends
of their bars.  Shorter bars should also have stronger knobby features.
We expect that the ratio of the $S_4$  and higher order 
components to the $S_2$ component of the surface brightness 
should depend upon Hubble type and bar length with larger values for earlier Hubble
types and shorter bars.  
Alternatively prominant knobby
ends may only develop in bars that are evolving slowly.
Then the ratio of the $S_4$ to the $S_2$ components
would anticorrelate with other phenomena associated with faster
evolution such as strong spiral structure, and assymetries in the
bar and spiral arms.  In this case strong knobby features are commonly seen in
early-type galaxies because their bars would be changing shape less
rapidly than those in late-type galaxies.
Bars which deviate from these patterns may be strongly influenced
by gas dynamics or have large velocity dispersions.

Because of the strong higher order components in early-type 
barred galaxies, all the higher order resonances are strong.
Vertical and planar stellar heating (increase in velocity dispersion) 
should be more efficient in these galaxies than in later type galaxies.
This suggests that the rate that bars evolve, disks thicken
and bulges grow could all be higher in earlier type barred galaxies than
in late-type ones.

\acknowledgments

I acknowledge many helpful sujestions and criticisms from E. Athanassoula.
I also acknowledge helpful discussions and correspondence 
with B. Elmegreen, D. Elmegreen and A. Gould. 
This paper was inspired by a talk given by D. Elmegreen at the
Barred Galaxy meeting, May 1995, in Tuscaloosa.
The OSU galaxy survey is being supported in part by NSF grant AST 92-17716.
OSIRIS was built with substantial aid from NSF grants AST 90-16112 and 
AST 92-18449.
A.C.Q. acknowledges the support of a Columbus fellowship and a 
grant for visitors from L'Observatoire de Marseille.

\clearpage

\begin{figure*}
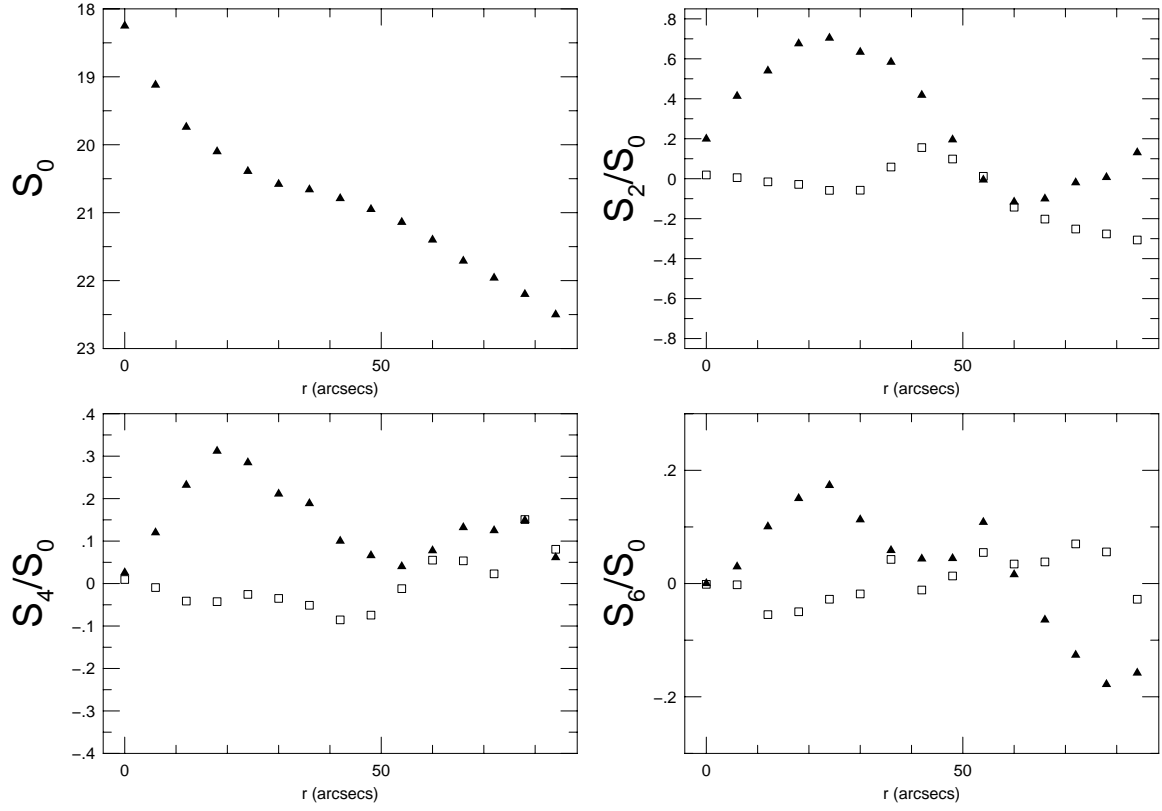

\caption[junk]{
The Fourier components of the $J$ surface brightness of NGC~1073.
$S_0$ is the azimuthally averaged component in mag/arcsec$^2$.
The $m=2,4$ and $6$ components are given divided by $S_0$.
For these components the solid
triangles show show the cosine components and open
squares show the sine components.
\label{fig:fig1} }
\end{figure*}

\begin{figure*}
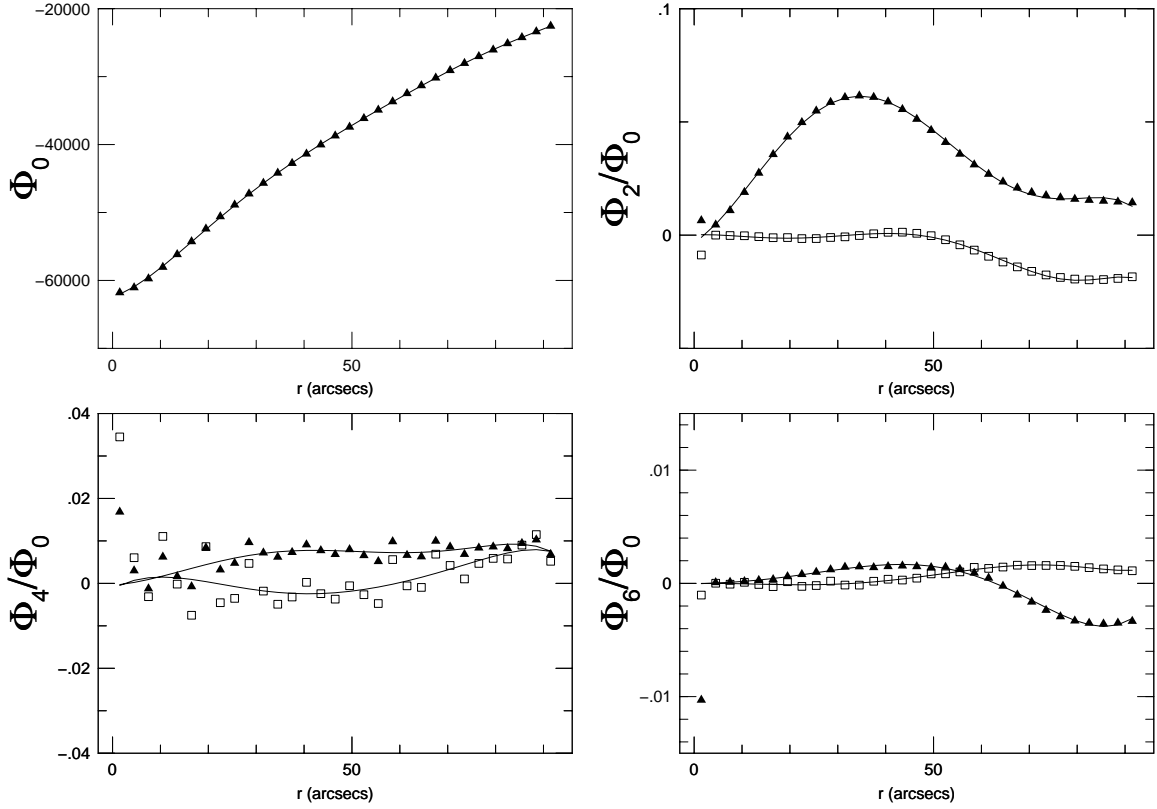

\caption[junk]{
The Fourier components of the potential of NGC 1073.
$\Phi_0$ is the azimuthally averaged component in (km/s)$^2$ for
distance and mass-to-light ratio given in the text.
The $m=2,4$ and $6$ components are given divided by $\Phi_0$.
For these components the solid
triangles show show the cosine components and open
squares show the sine components.
Polynomial fits to these functions are shown as solid lines.
\label{fig:fig2} }
\end{figure*}

\begin{figure*}
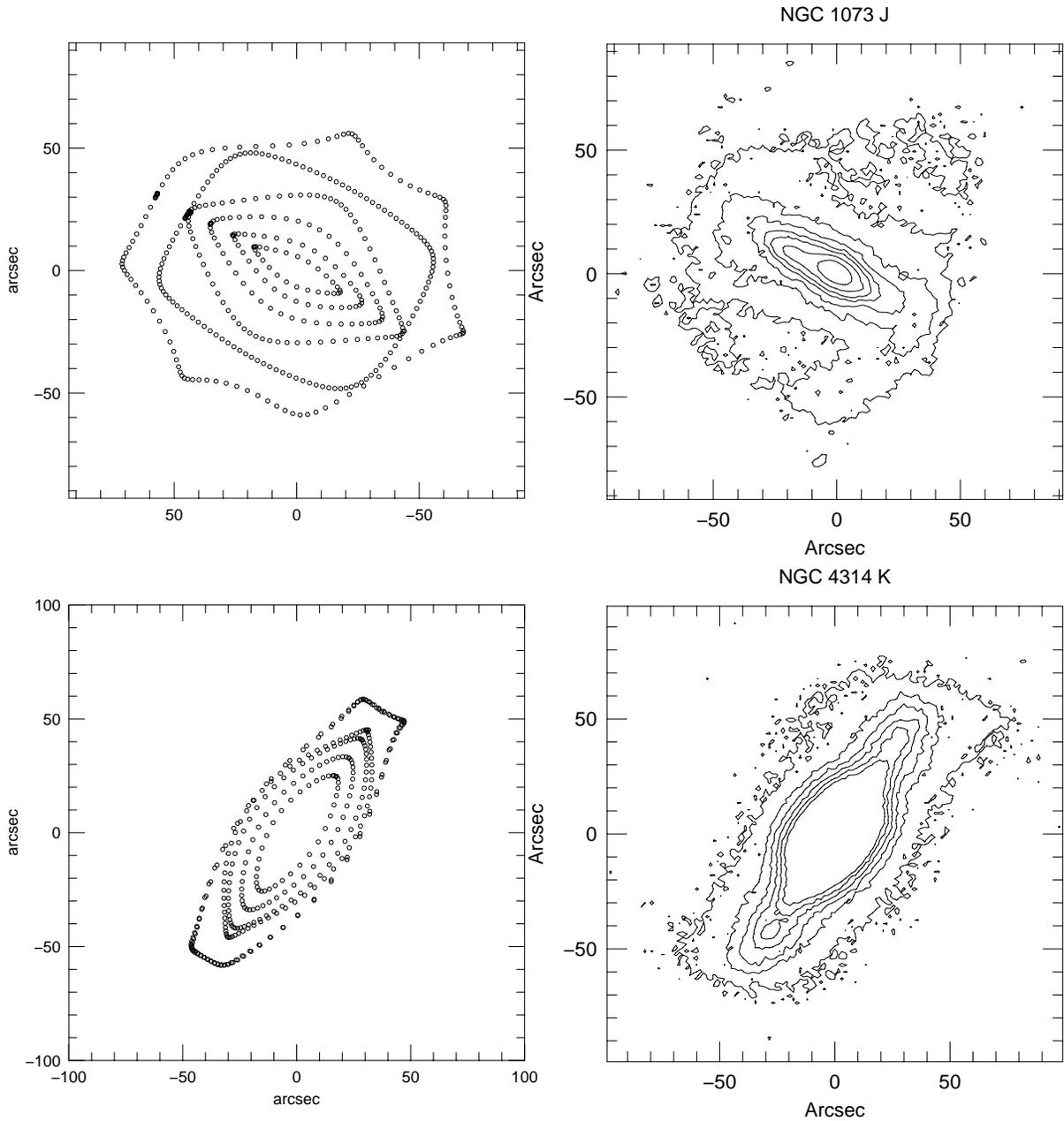

\caption[junk]{
a) Prograde stellar orbits in NGC~1073 for a corotation radius at
$r=85''$ in the frame in which the bar is stationary.
Points are plotted at equal time intervals in a single orbit.
The density of the galaxy should increase where the velocity decreases.
b) Contour plot of NGC~1073 of $J$ surface brightness.
The lowest contour is at a level 21.13 mag/''$^2$.
The difference between the first five contours is equal to the surface
brightness which corresponds to the magnitude of the lowest contour.
c) Prograde stellar orbits in NGC~4314 for a corotation radius at $r=70''$.
d) Contour plot of NGC~4314 of $K$ surface brightness.
The lowest contour is at a level of $19.45$ magnitudes per arcsec$^2$.
The difference between contours is equal to the surface
brightness which corresponds to the magnitude of the lowest contour.
From Quillen et al. (1994).
\label{fig:fig3ab} }
\end{figure*}

\begin{figure*}
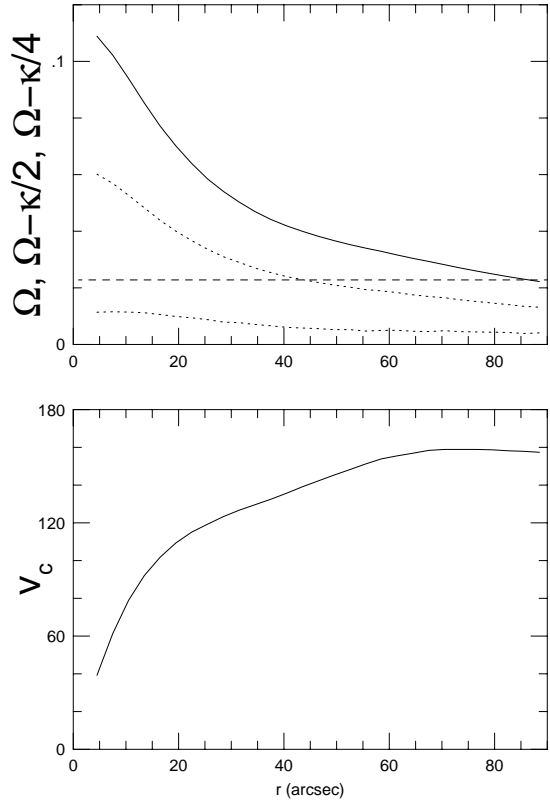

\caption[junk]{
a) Angular rotation rate for NGC~1073 $\Omega=v_c/r$ for $v_c$
the circular rotation speed is plotted as the 
solid line and is derived from the azimuthally symmetric component 
of the potential $r \Omega^2= d\Phi_0/dr$.  $\Omega$ is given in Myr$^{-1}$
for the mass-to-light ratio and distance given in the text.
As an aid to finding the resonances, we have also plotted
$\Omega - \kappa/2$ and $\Omega - \kappa/4$ as dotted lines
where $\kappa$ is the epicyclic frequency.
For a bar angular rotation rate (shown as a horizontal dashed line)
such that the corotation radius is
at $r=85''$, the inner 4:1 Lindblad
resonance is at $r \sim 45''$ and there is no Inner Lindblad Resonance.
b) Circular rotation speed $v_c$ in km/s.
\label{fig:fig4} }
\end{figure*}

\begin{figure*}
\caption[junk]{
Non-periodic orbits in NGC~1073.  Each column shows orbits 
with the same initial position as the periodic orbit
shown in the bottom panel but with varying initial tangential velocities.
The initial radius is a) $30''$ and b) $50''$.
\label{fig:fig5} }
\end{figure*}

\begin{figure*}
\caption[junk]{
Non-periodic orbits in NGC~4314.  Each column shows orbits 
with the same initial position as the periodic orbit
shown in the bottom panel but with varying initial tangential velocities.
The initial radius is a) $40''$, b) $50''$ and c) $65''$.
\label{fig:fig6} }
\end{figure*}

\begin{figure*}
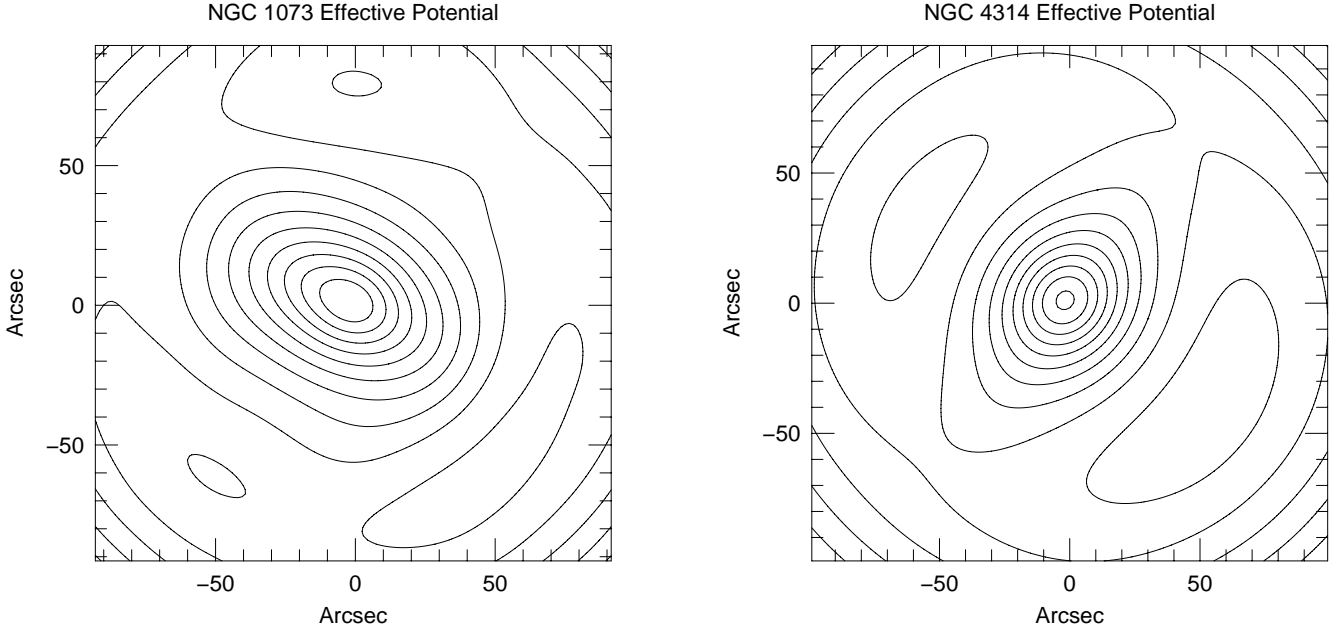

\caption[junk]{
Effective potentials for a) NGC~1073 and b) NGC~4314 for bar pattern
speeds given in the text.   The difference between contour levels is
$2.5 \times 10^3$ (km/s)$^2$ for NGC~1073 and twice this for NGC~4314.
Since the corotation radius for NGC~1073 is much greater than its bar length
the potential is rounder near the Lagrange radius
than in NGC~4314.   Near the Lagrange radius in NGC~1073 structure from the
spiral arms can be seen (compare to Figure 3b).
\label{fig:fig7} }
\end{figure*}
\vfill

\clearpage

\begin{deluxetable}{crrrrrrrr}
\footnotesize
\tablewidth{300pt}
\tablecaption{Properties of Periodic Orbits \label{tbl-1}}
\tablehead{ \colhead{Galaxy} & \colhead{$r_{max}$}   & \colhead{$r_{max} \over r_{min}$} & \colhead{$v_{max}\over v_{min}$}   & \colhead{Curvature$_{max}$}} 
\startdata
NGC~4314  & $30''$ & 2.94 & 6.22 & 12.9    \cr
NGC~4314  & $40''$ & 2.66 & 5.55 & 14.3    \cr
NGC~4314  & $50''$ & 2.50 & 6.79 & 20.9    \cr
NGC~4314  & $55''$ & 2.42 & 10.7 & 51.9    \cr
          &      &      &      &         \cr
NGC~1073  & $20''$ & 3.49 & 10.2 & 22.2   \cr
NGC~1073  & $30''$ & 2.80 & 7.84 & 17.4   \cr
NGC~1073  & $40''$ & 2.14 & 4.69 & 19.8   \cr
NGC~1073  & $50''$ & 1.70 & 5.25 & 26.9   \cr
%
%
\enddata
%
%
\end{deluxetable}

\end{document}